\documentclass[twocolumn,showpacs,preprintnumbers,amsmath,amssymb,pre,
showkeys,groupedaddress,superscriptaddress]{revtex4}
\usepackage{graphicx}% Include figure files
\usepackage{dcolumn}% Align table columns on decimal point
\usepackage{bm}% bold math

\begin{document}

%\preprint{APS/123-QED}

\title{Analytical kinetics of clustering processes with cooperative \\
      action of aggregation and fragmentation}

\author{Vladimir M. Dubovik}
\affiliation{Bogoliubov Laboratory of Theoretical Physics (BLTP), JINR}
\email{dubovik@thsun1.jinr.ru}
\author{Arkadi G. Galperin}
\affiliation{Laboratory of High Energies (LHE), JINR}
\email{galperin@vxjinr.jinr.ru}
\author{Viktor S. Richvitsky}
\affiliation{Laboratory of Information Technologies (LIT), JINR}
\email{rqvtsk@cv.jinr.ru}
\collaboration{Joint Institute for Nuclear Research (JINR), Dubna,
Moscow reg., 141980, Russia}
\author{Aleksey A. Lushnikov}
 \affiliation{ Karpov Institute of Physical Chemistry (NIFHI), Moscow}
\email{alush@cc.nifhi.ac.ru}
%\collaboration{ Karpov Institute of Physical Chemistry (NIFHI), Moscow}

\date{\today}% It is always \today, today,
             %  but any date may be explicitly specified

\begin{abstract}
Some models of clustering processes are formulated and analytically solved
employing generating functions methods. Those models include events which
result from combined action of the coagulation and fragmentation processes.
Fragmentation processes of two kinds, so-called similar and arbitrary, ones,
are brought forward, and the explicit forms of their solutions are produced.
This implies some possibility of existence of different aggregation mechanisms
for clusters creation differing in their inner structure. All the models are
based on "the three-level bunch" scheme of interaction between the system
states. Those states are described in terms of the probability to find the
system in the state with an exactly given number of clusters. The models are
linear in the probability functions due to the assumption that the rates of
elementary acts are permanent.
Some peculiarities of application of the generating function method to
solution of the linear differential-difference equations are revealed.
The illustration of the problem in terms of a traffic jam picture is not a
specific one.
\end{abstract}

%PACS, the Physics and Astronomy Classification Scheme.


\maketitle

\section{\label{intr}Introduction}
In the present work we develop and define more accurately our earlier
considerations \cite{DGLR}.
A cluster is generally understood as either a number of things of the same kind growing
together or number of particles, objects, etc., in a small, close group.
This idea is a very general and not a specified one. An aggregation process
in the form $au+bu=(a+b)u$ is referred as coagulation, when coefficients
$a,b$ represent the quantities of the scale unit of $u$ which are coalesced
with time. Aggregation and fragmentation are a couple of mutually inverse
processes. The physical scales ( spatial, temporal, value of masses, e.t.c.)
vary in the many orders of a magnitude for such a processes. That is why, an
idea of existence of a universal description of the above phenomena arises
and the unification of both direct and inverse processes in a general class
of clustering processes seems being of a natural one.

The clustering processes resemble to a certain degree the nucleation
processes. This roots in the mathematical description being general for
kinetics of such processes despite the lack of obvious resemblance in their
actual more precise details. On the one hand, we can see those approaches in
the course of investigations on the molecular and submolecular level, in
theories of condensed matter, nuclei and nuclear chains \cite{Nuconf}. On the
other hand, clustering of disperse systems are considered in astrophysics
(forming of cosmic objects), atmospheric science, chemistry, ... \cite{Astro}.

For example,
an expanding universe is formed not at once. Clusters grow by coalescence of
smaller clusters. Their growth kinetics is like to the kinetics of
coagulation. In what follows we formulate basic equations and outline the
methods for their solution. Moreover, one could expect that theoretical tools
developed to describe physical systems can be exploited in other fields, such
as ecology of computation \cite{Huberman}
or biology, economics, transport problem, etc. \cite{HH,NicolPrig}.

Consider the kinetics of formation of a $G$-cluster using the picture of a
one-way motor lane. We assume that the starting configuration is $G$
independent cars on the motor lane, the leading one being the slowest, and no
one can pass over each other. So, each initial cluster contains one car.

The process begins at $t=0$. On passing some time $t$, the initial $G$ cars
group in clusters containing $g_1, g_2\ldots g_s$ cars. These clusters go on
to coalesce. The problem is to determine the time evolution of the probability $w_s(g_1,g_2\ldots g_s;t)$
to find  $s$ clusters $g_1, g_2\ldots g_s$. The sum of their masses $g_k$
are subjected to the constraint (conservation law):
\begin{equation}\begin{array}{c}
\sum\limits_{k=1}^s g_k=G
\end{array}\label{intr1}\end{equation}
in the system. Thus, we concern only of non-relativistic events and study
systems with a permanent (additive) mass and a finite number of particles.

Our goal is to formulate and investigate the exactly solved models of the
clustering (dissociation) processes, including those which result from
combined action of a certain aggregation and fragmentation.

The paper is organized as follows. In Section~\ref{NC}, we explain the
stochastic motion of our objects, obtain the probability of finding out the
system in the state of exactly $s$ clusters and dependent on the time average
number of clusters by means of introducing a generating function. Later on,
we formulate a master equation governing the time evolution of the probability
of finding the clusters of various masses (Sect.~\ref{MDPAP}). We solve that
problem applying the Laplace transformation with respect to $t$ to master
equation (Sect.~\ref{SBE}). Then we find the probability to detect a cluster
of assigned mass $g$ by summation of $w_s$ over all $g_k$ irrespective of the
distribution of other participants, except for the selected one
(Sect.~\ref{SCDPAP}). Some properties of those convolutions are true due to
the isomorphism between a set of generating functions with a product
operation and a set of $w(g)$ with convolution (see Appendix as well),
which have the structures of semigroups with unit. In the following we
consider processes of similar fragmentation (Sect.~\ref{SFP}), combined
action of aggregation and similar fragmentation (Sect.~\ref{PASF}), and the
process with an arbitrary fragmentation (Sect.~\ref{PARF}). In conclusion, we
discuss our results.

\section{\label{NC}Number of Clusters: Pure Aggregation Process}

Let $\gamma$ be the rate of an elementary coalescence act; two adjacent
clusters produce a single one (for instance, a dimer is formed when a car
catches up another one). We assume that $\gamma$ is $g$--independent. Then we
can characterize the situation by the number of intervals between adjacent
clusters. If there are $s$ clusters in the system, the number of intervals is
$s-1$. Each coalescence act annihilates one interval. The number of ways to
do this is exactly equal to the number of intervals.

Let $W(s,t)$ be the probability of meeting exactly $s$ clusters at
the time $t$. Then:
\begin{equation}\begin{array}{c}
\frac{dW(s,t)}{dt}=\gamma
[sW(s+1,t)-(s-1)W(s,t)] .
\end{array}\label{nc1}\end{equation}
One can observe a "three-level bunch" scheme of transitions between the
three nearest (on $s$) states of the system under consideration.
Equation (\ref{nc1}) should be supplemented with the initial conditions:
\begin{equation}\begin{array}{c}
W(s,0)=W_0(s) .
\end{array}\label{nc2}\end{equation}
In particular, if initially there were exactly $G$ independent cars, the
function $W_0(s)$ obeys the equation
\begin{equation}\begin{array}{c}
W_0(s)=\Delta(s-G) ,
\end{array}\label{nc3}\end{equation}
with $\Delta$ being the Kronecker delta: $\Delta(0)=1$, and $\Delta=0$
otherwise.

Equation (\ref{nc1}) can be solved by introducing the generating function:
\begin{equation}\begin{array}{c}
F(z,t)=\sum\limits_sW(s,t)z^{s-1} .
\end{array}\label{nc4}\end{equation}
Combining Eqs. (\ref{nc1}) and (\ref{nc4}) gives:
\begin{equation}\begin{array}{c}
\partial_t F=(1-z)\partial_zF .
\end{array}\label{nc5}\end{equation}
The rate $\gamma$ is included into the definition of time. The initial
condition for an initially monodisperse system is rewritten in terms of $z$ as:
\begin{equation}\begin{array}{c}
F(z,0)=z^{G-1} .
\end{array}\label{nc6}\end{equation}
The solution of Eq. (\ref{nc5}) with the initial condition, Eq. (\ref{nc6}), has the form:
\begin{equation}\begin{array}{c}
F(z,t)=[1-e^{-t}(1-z)]^{G-1} .
\end{array}\label{nc7}\end{equation}
The probability $W(s,t)$ is thus expressed in terms of binomial distributions:
\begin{equation}\begin{array}{c}
W(s,t)=C^{s-1}_{G-1}e^{-(s-1)t}(1-e^{-t})^{G-s} .
\end{array}\label{nc8}\end{equation}
It is no problem to find the time dependence of the average number of clusters:
\begin{equation}\begin{array}{c}
\bar s(t)=\partial_zzF(z,t)|_{z=1}=1+(G-1)e^{-t} .
\end{array}\label{nc9}\end{equation}

\section{\label{MDPAP}Mass Distribution in a Pure Aggregation Process}

In analogy with the kinetics of disperse systems, we shall refer to $g_k$ as
the cluster mass. Our goal now is to formulate the master equation
governing the time evolution of the probability $w_s(g_1,g_2 \ldots g_s;t)$
to find the clusters of masses $g_1, g_2\ldots$ at the time $t$. This equation
is formulated as follows:
% below there is a trick to divide a long expression into N rows: all these
% lines express the only one mathematical formula
\begin{equation}\begin{array}{c}
\frac{dw_s}{dt}=
\sum\limits_{[g'],k}w_{s+1}(g_1\ldots g'_k,g'_{k+1},g'_{k+2}\ldots g'_{s+1}) \\
\times\Delta(g_k-g'_k-g'_{k+1}) \\
\times\Delta(g'_{k+2}-g_{k+1})\ldots\Delta(g'_{s+1}-g_s)-(s-1)w_s .
\end{array}\label{MDPAP1}\end{equation}
The meaning of the terms in the r.h.s. of Eq. (\ref{MDPAP1}) is rather apparent. The
rate of losses is simply proportional to the number of empty intervals
(the rate constant $\gamma$ is included into the definition of time). The
gain occurs each time when two clusters of masses $g'_k$ and $g'_{k+1}$
coalesce
producing a new cluster of mass $g_k$. Other $\Delta$-s simply restore
the serial numbers of $g_i$ clusters with $ i<k $ for the system of $s$
clusters.

Of course, initial conditions to Eq. (\ref{MDPAP1}) should also be specified. We
again assume that initially there were G separate cars:
\begin{equation}\begin{array}{c}
w_G(1,1\ldots 1, t=0)=1 ,
\end{array}\label{MDPAP2}\end{equation}
and all other probabilities are $0$.

\section{\label{SBE}Solution to Basic Equation}

On applying the Laplace transformation with respect to $t$ gives, instead of
Eq. (\ref{MDPAP1}),
\begin{equation}\begin{array}{c}
(p+s-1)\bar w_s(g_1,g_2\ldots) \\
=\sum\limits_{[g'],k}\bar
w_{s+1}(g_1\ldots g'_k,g'_{k+1},g'_{k+2}\ldots g'_{s+1}) \\
\times\Delta(g_k-g'_k-g'_{k+1}) \\
\times\Delta(g'_{k+2}-g_{k+1})\ldots\Delta(g'_{s+1}-g_s) ,
\end{array}\label{SBE1}\end{equation}
where barred $\bar w$ stands for the Laplace transform of $w(g_1,g_2\ldots;t)$.
The last equation of this set is readily solved (Eq. (\ref{nc8})) to yield:
\begin{equation}\begin{array}{c}
\bar w_G=\frac1{p+G-1} .
\end{array}\label{SBE2}\end{equation}
Now let us try to seek for a solution to Eq. (\ref{MDPAP1}) in the form:
\begin{equation}\begin{array}{c}
\bar w_s(g_1,g_2\ldots ;,p)=\frac{A_s(g_1,g_2\ldots)}
{(p+G-1)(p+G-2)\ldots(p+s-1)}  ,
\end{array}\label{SBE3}\end{equation}
where the coefficients $A$ are independent of $p$ and satisfy the following
set of recurrence relations:
\begin{equation}\begin{array}{c}
A_s(g_1,g_2\ldots) \\
=\sum\limits_{[g'],k}A_{s+1}(g_1\ldots g'_k,g'_{k+1},g'_{k+2}\ldots g'_{s+1}) \\
\times\Delta(g_k-g'_k-g'_{k+1}) \\
\times\Delta(g'_{k+2}-g_{k+1})\ldots\Delta(g'_{s+1}-g_s) .
\end{array}\label{SBE4}\end{equation}
A useful sum rule
\begin{equation}\begin{array}{c}
Q_s=sQ_{s+1},
\end{array}\label{SBE41}\end{equation}
follows immediately from Eq. (\ref{SBE4}),
where $Q_s=\sum A_s(g_1,g_2\ldots g_s)$ (summation runs over all $g$), or
\begin{equation}\begin{array}{c}
Q_s=\frac{(G-1)!}{(s-1)!} .
\end{array}\label{SBE42}\end{equation}
In fact, the expression $A_s(g_1,\ldots ,g_s)$ depends on $s$ only. It does
not depend on the distribution of numbers $g_1,\ldots ,g_s$ provided that
$\sum_{k=1}^s{g_k}=G,\ \ \ g_k\ge 1$ are conserved.

It can be seen from Eqs. (\ref{MDPAP1}), (\ref{SBE1}), (\ref{nc8}), and (\ref{nc1}) that the problem under
consideration splits into two subproblems. The first one is the time
evolution problem. It deals with transitions between different states of the
system and connects each three nearest adjacent states. The second subproblem
is to scrutinize mass spectra. It is a pure combinatorial task. In fact, we
have to deal with some population dynamics. Mass spectra at instant $t$
originate from the interchange of generations at a given $G$, and the proper
weights depend on the whole set of possible transitions from $s+1$-states to
the $s$-state under consideration.

%\ins{fig1}{dbw.eps}{6in}{3in}%{h}
   \begin{figure*}%[h]
%    \centering
    \includegraphics[width=7in,totalheight=3in]{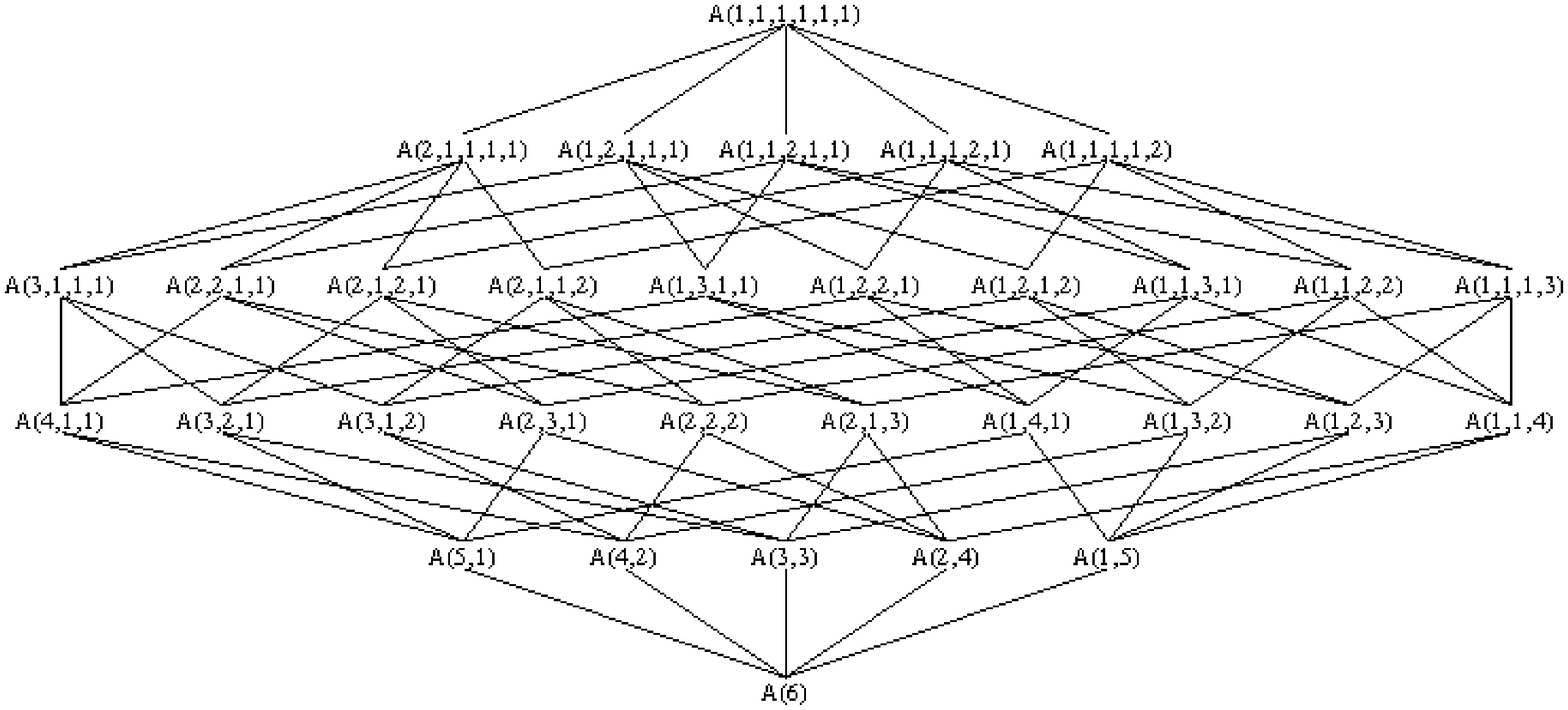}
    \caption{Generation scheme for G=6}
   \end{figure*}

With the induction method, one obtains that:
\begin{itemize}
\item $A_G(1,\ldots ,1)=1$ is the only possible value;
\item the number of terms in formula Eq. (\ref{SBE4}) is equal to
$$\sum_{k=1}^s{g_k-1}= \sum_{k=1}^s{g_k}-\sum_{k=1}^s 1 = G-s$$
for each $s$ fixed and $k$ fixed. Under the inductive assumption, one can
write down $A_{s+1}(g'_1,\ldots ,g'_{s+1})=A_{s+1}$. From this it follows
that $A_s(g_1,\ldots ,g_s)=(G-s)A_{s+1}$ irrespective of a specific
distribution of $g_1,\ldots ,g_s$.
\end{itemize}

The recurrence equations obtained just now
\begin{equation}\begin{array}{c}
A_s=(G-s)A_{s+1},\\
A_G=1
\end{array}\label{SBE43}\end{equation}
have solutions
\begin{equation}\begin{array}{c}
 A_s=(G-s)! .
\end{array}\label{SBE5}\end{equation}

The time dependence can be readily restored by using the inversion
\begin{equation}\begin{array}{c}
\frac1{(p+s-1)(p+s)\ldots (p+G-1)} {\longrightarrow} \\ \\
{\longrightarrow} \frac1{(G-s)!}e^{-(s-1)t}(1-e^{-t})^
{G-s} .
\end{array}\label{SBE6}\end{equation}
Equations (\ref{SBE4}),  (\ref{SBE41}), and (\ref{SBE6}) reproduce Eq.
(\ref{nc8}) as well.

The final result is formulated as follows:
\begin{equation}\begin{array}{c}
w_s(g_1,g_2,\ldots g_s;t)=e^{-(s-1)t}(1-e^{-t})^{G-s} \\
\times\Delta(G-g_1-g_2-\ldots g_s) .
\end{array}\label{SBE7}\end{equation}

\section{\label{SCDPAP}Single-cluster Distribution in a Pure Aggregation Process}

Here we determine the probability to find a cluster of mass $g$ irrespective
of the distribution of other participants. To this end we sum
$w_s$ over all $g_k$ except one ($g_1$, for example):
\begin{equation}\begin{array}{c}
w(g,t)=\sum\limits_{g_k}w_s(g,g_2,\ldots g_s;t)  \\
=e^{-(s-1)t}(1-e^{-t})^{G-s} \\
\times\sum\limits_{g_k}\Delta(G-g-g_2-\ldots g_s) .
\end{array}\label{SCDPAP1}\end{equation}
Using the identities
\begin{equation}\begin{array}{c}
\Delta(q)=\left\{\begin{array}{l}{1,\ q=0}\\{0,\ q=1,2,...}\end{array}\right\}
=\frac1{2\pi i}\oint\frac{dz}{z^qz},\\ \\
\left\{\begin{array}{l}{0,\ q=0}\\{1,\ q=1,2,...}\end{array}\right\}
=\frac1{2\pi i}\oint\frac{zdz}{z^q(1-z)}, \\ \\
\frac1{2\pi i}\oint\frac{dz}{z^{r+1}(1-z)^{R+1}}=C_{R+r}^r=C_{R+r}^R ,
\end{array}\label{SCDPAP2}\end{equation}
one finds the convolution in Eq. (\ref{SCDPAP1}):
\begin{equation}\begin{array}{c}
w(g,t)=e^{-(s-1)t}(1-e^{-t})^{G-s} \\
\times\frac1{2\pi i}\oint\frac{z^{s-1}dz}{z^{G-g}(1-z)^{s-1}z} \\
=C_{G-g-1}^{s-2}e^{-(s-1)t}(1-e^{-t})^{G-s} .
\end{array}\label{SCDPAP3}\end{equation}
Some important properties of the convolutions are discussed in Appendix.

%\ins{fig2}{cbw.eps}{4in}{2in}%{h}
   \begin{figure}[h]
%    \centering
    \includegraphics[width=3in,totalheight=2in]{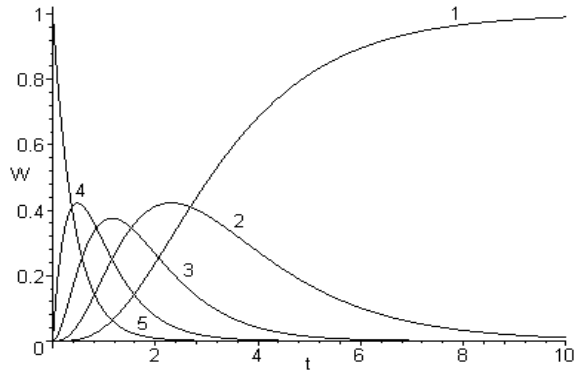}
    \caption{W-functions versus time for pure aggregation process. $G=5, \gamma=1$.
    All curves $W(s,t)$ are marked by the number $s$ of the proper state.}
   \end{figure}

\section{\label{SFP}Similar Fragmentation Process }

Let us consider a process of pure fragmentation (dissociation, decay) of
clusters.
We assume the inner $G$-cluster structure at $t=0$ to be similar to the
picture at $t=0$ with $G$ cars of unit mass in Sect.~\ref{NC} and, thus, with
$G-1$ intervals between them . An analogous assumption applies to all the
$s$-clusters at $t\ne 0$. We will understand "the similarity of an inner
cluster structure and an outer one" in such a purport. In addition, let us
suppose a qualitative equivalence of all the inner intervals between
constituents of a cluster.

At the above pure aggregation mechanism the place where a coalescence act has
happened is partly forgotten. When two adjacent clusters enumerated $k_s$ and
$(k+1)_s$ according to enumeration of an $s$-cluster state with sizes (see
below) $(g\prime_k)_s$ and $(g\prime_{k+1})_s$ coagulate, a new cluster of
the size $(g_k)_{s-1}$=$(g\prime_k)_s+(g\prime_{k+1})_s$ arises. The cluster
ordinal number $k_{s-1}$ is given in terms of the new $(s-1)$-cluster state
which originates from the act of coagulation as in Sect.~\ref{MDPAP}.

Evidently, a partial loss of memory on the way in which the microscopic
state has been created comes about at the very place because the inverse
problem of one-to-one rebuilding of the
previous $s$-cluster state (the outer state) cannot be solved. On the other
hand, the things are the same with the interior structure of some cluster.
Its cluster parents cannot be reconstructed one-to-one as well. When one looks
at a system state or some cluster as something given, the exact information
about ordinal numbers and sizes of adjacent cluster parents and even the
mother state is forgotten. That is why one can talk about a loss of memory in
such a process. In the whole it is the reason to postulate some equivalence
of constituents and intervals between them in the inner cluster space.

This set of assumptions results in the dependence of the probability functions
on the size of a cluster and the time only.

Let us realize the cluster size as the number of particles confined into the
given cluster, let $\gamma$ be the rate of an elementary fragmentation act.

If a fragmentation rate is proportional to the cluster size minus unit, i.e.,
the number of possible rupture places is equal to the number of inner
intervals, the equation
\begin{equation}\begin{array}{c}
\frac{dW(s,t)}{dt}= \gamma [(G-s+1)W(s-1,t)-(G-s)W(s,t)] ,
\end{array}\label{SFP1}\end{equation}
with initial conditions
\begin{equation}\begin{array}{c}
W(1,0)=1; \ \ \ \ W(s,0)=0, \ \ \ if \ s\ne 1 ,
\end{array}\label{SFP2}\end{equation}
describes the process under consideration. The r.h.s. of Eq. (\ref{SFP1}) consists
of a gain term
due to decay of clusters belonging to an $s-1$-cluster state and a loss term
due to decays of those clusters belonging to $s$-cluster state that produces
the clusters pertaining to an $s+1$-cluster state.
Equation (\ref{SFP1}) can be solved by using the generating function introduced by the
equation
\begin{equation}\begin{array}{c}
{\frac {\partial }{\partial t}}\,{\rm F}(z, \,t) - ( - z^{2}\,
\gamma  + z\,\gamma )\,{\frac {\partial }{\partial z}}\,{\rm F}(
z, \,t) \\
=\gamma \,(z\,G - z - G + 1)\,{\rm F}(z, \,t) ,
\end{array}\label{SFP3}\end{equation}
with the solution
\begin{equation}\begin{array}{c}
F(z,t)=
 - { \frac {(z + e^{( - \gamma \,t)} - e^{( - \gamma \,t)}\,z)^{G}}
              { - z - e^{( - \gamma \,t)} + e^{( - \gamma\,t)}\,z}} .
\end{array}\label{SFP4}\end{equation}
Of course, one recognizes a usual Poissonian process here. It seems
contextual and, hence, quite reasonable to name such a process as a similar
fragmentation (process).

For example, for $G=5$
\begin{equation}\begin{array}{c}
{{\rm W}_{1}}=e^{( - 4\,\gamma \,t)} ,
\end{array}\label{SFP5}\end{equation}
\begin{equation}\begin{array}{c}
{{\rm W}_{2}}=4\,e^{( - 3\,\gamma \,t)} - 4\,e^{( - 4\,\gamma \,
t)} ,
\end{array}\label{SFP6}\end{equation}
\begin{equation}\begin{array}{c}
{{\rm W}_{3}}=6\,e^{( - 2\,\gamma \,t)} - 12\,e^{( - 3\,\gamma
\,t)} + 6\,e^{( - 4\,\gamma \,t)} ,
\end{array}\label{SFP7}\end{equation}
\begin{equation}\begin{array}{c}
{{\rm W}_{4}}=4\,e^{( - \gamma \,t)} - 12\,e^{( - 2\,\gamma \,t)
} + 12\,e^{( - 3\,\gamma \,t)} - 4\,e^{( - 4\,\gamma \,t)} ,
\end{array}\label{SFP8}\end{equation}
\begin{equation}\begin{array}{c}
{{\rm W}_{5}}=1 - 4\,e^{( - \gamma \,t)} + 6\,e^{( - 2\,\gamma
\,t)} - 4\,e^{( - 3\,\gamma \,t)} + e^{( - 4\,\gamma \,t)} .
\end{array}\label{SFP9}\end{equation}
The average number of clusters reads
\begin{equation}\begin{array}{c}
 \bar s(t)=\partial_zzF(z,t)|_{z=1}=
e^{( - \gamma \,t)} + G\,(1 - e^{( - \gamma \,t)}) .
\end{array}\label{SFP10}\end{equation}

\section{\label{PASF}Process of Aggregation and Similar Fragmentation }

Let us consider such a clustering process, which runs as a result of some
combined action both of the aggregation and the similar fragmentation.
Let $\gamma_1$ and $\gamma_2$ be constant rates of an elementary coalescence
act and an elementary fragmentation act, respectively,
\begin{equation}\begin{array}{c}
\frac{dW(s,t)}{dt} \\
=\gamma_1 [sW(s+1,t)-(s-1)W(s,t)] \\
-\gamma_2 [(G-s)W(s,t)-(G-s+1)W(s-1,t)] ,
\end{array}\label{PASF1}\end{equation}
with initial conditions
\begin{equation}\begin{array}{c}
W(G,0)=1; \ \ \ \ W(s,0)=0, \ \ \ if \ s\ne G .
\end{array}\label{PASF2}\end{equation}
The r.h.s. of Eq. (\ref{PASF1}) consists of gain terms due to coagulation of
clusters from an $(s+1)$-cluster state and dissociation of those clusters
belonging to an $(s-1)$-cluster state and loss terms due to simultaneous
coalescence and dissociation of clusters belonging to an $s$-cluster state.
To make things more clear, we could rewrite Eq. (\ref{PASF1}) in the form
\begin{equation}\begin{array}{c}
\frac{dW(s,t)}{dt} \\
=(\gamma_1 {sW(s+1,t}) + \gamma_2 {(G-s+1)W(s-1,t)}) \\
-(\gamma_1 (s-1)W(s,t) + \gamma_2 (G-s)W(s,t)).
\end{array}\label{PASF1'}\end{equation}

These equations can be solved by using the generating function $F(z,t)$
defined by the equation

\begin{equation}\begin{array}{c}
{\frac{\partial }{\partial t}}\,{\rm F}(z, \,t) \\
- ( - z^{2}\,
{ \gamma _{2}} - z\,{\gamma _{1}} + z\,{ \gamma _{2}} + {\gamma _{1}})\,
{\frac{\partial }{\partial z}}\,{\rm F}(z, \,t) \\
={ \gamma_{2}}\,(z\,G - z - G + 1)\,{\rm F}(z, \,t) ,
\end{array}\label{PASF3}\end{equation}

whose solution is
\begin{equation}\begin{array}{c}
F(z,t) \\
={{\rm \gamma}\,({
\frac {z\,{\gamma _{2}} + {\gamma _{1}}\,e^{( - t\,{\rm \gamma})}\,z
 - {\gamma _{1}}\,e^{( - t\,{\rm \gamma})} + {\gamma _{1}}}{{\rm \gamma
}}} )^{G}}/ \\
({z\,{\gamma _{2}} + {\gamma _{1}}\,e^{( - t\,{\rm \gamma})
}\,z - {\gamma _{1}}\,e^{( - t\,{\rm \gamma})} + {\gamma _{1}}}) ,
\end{array}\label{PASF4}\end{equation}
where
\begin{equation}\begin{array}{c}
{\rm \gamma} = {\gamma _{1}} + {\gamma _{2}} .
\end{array}\label{PASF5}\end{equation}

For example, for $G=5$

\begin{equation}\begin{array}{c}
{{\rm W}_{1}}=.6830134554\,(e^{( - 1.1\,t)} - 1)^{4},
\end{array}\label{PASF6}\end{equation}

\begin{equation}\begin{array}{c}
{{\rm W}_{2}}= - 2.732053822\,(e^{( - 1.1\,t)} - 1)^{3}\,(.1 + e
^{( - 1.1\,t)}),
\end{array}\label{PASF7}\end{equation}

\begin{equation}\begin{array}{c}
{{\rm W}_{3}}=4.098080732\,(e^{( - 1.1\,t)} - 1)^{2}\,(.1 + e^{(
 - 1.1\,t)})^{2},
\end{array}\label{PASF8}\end{equation}

\begin{equation}\begin{array}{c}
{{\rm W}_{4}}= - 2.732053822\,(e^{( - 1.1\,t)} - 1)\,(.1 + e^{(
 - 1.1\,t)})^{3},
\end{array}\label{PASF9}\end{equation}

\begin{equation}\begin{array}{c}
{{\rm W}_{5}}=.6830134554\,(.1 + e^{( - 1.1\,t)})^{4}.
\end{array}\label{PASF10}\end{equation}

The average number of clusters is given by the formula
\begin{equation}\begin{array}{c}
 \bar s(t)=\partial_zzF(z,t)|_{z=1} \\
 = {1 - {\frac {{\gamma _{2}} + {\gamma _{1}}
\,e^{( - t\,{\rm \gamma})}}{{\rm \gamma}}}  + {\frac {G\,
({\gamma _{2}} + {\gamma _{1}}\,e^{( - t\,{\rm \gamma})})}{{\rm \gamma}
}} }.
\end{array}\label{PASF11}\end{equation}

%\ins{fig3}{abw.eps}{4in}{2in}%{h}
   \begin{figure}[h]
%    \centering
    \includegraphics[width=3in,totalheight=2in]{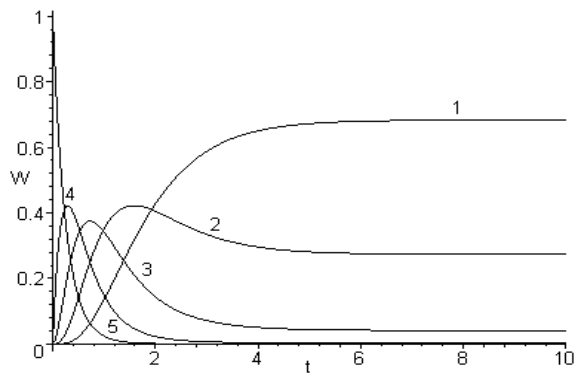}
    \caption{W-functions. Process of aggregation and similar fragmentation. $G=5, \gamma =1.5, \gamma_1=1, \gamma_2=0.5$.
    All curves $W(s,t)$ are marked by the number $s$ of the proper state.}
   \end{figure}

\section{\label{PARF}Process with an Arbitrary Fragmentation }

Let us assume that clusters lose the memory of their aggregation history
completely. This means that a hypothetical mechanism of such a kind
aggregation loses not only the place of a coalescence act (Sect.~\ref{SFP}),
but in this case the inner structure of a cluster does not coincide with
outer one, becomes a uniform one, and let us say "is beside itself". Shortly
speaking, the mass distribution inside a cluster becomes a uniform one under
such an aggregation, i.e., the inner structure does not depend on the cluster
size (see  Sect.~\ref{SFP}, and hence, the actual number of inner intervals
(see Sect.~\ref{PASF}) is equal to $0$.

However, let us apply the "three-level bunch" scheme of transitions
(connections) between the three nearest cluster-states, widely used above.
That is, in fact, one of the main subjects under consideration in this paper.

In this case, it is reasonable to assume that the fragmentation rate depends
solely on the outer cluster state indices. I.e., the rate depends on the
integer number of the clusters or the intervals minus unit between them
for the states belonging to such a three-level bunch. Adding the contextual
"natural" condition of stopping the fragmentation process at $s=G$,
the corresponding equation reads
\begin{equation}\begin{array}{r}
\frac{dW(s,t)}{dt}=\gamma_1 [sW(s+1,t)-(s-1)W(s,t)] - \gamma_2 \\
\times[(1-\Delta(G-s))sW(s,t)-(s-1)W(s-1,t)] ,
\end{array}\label{PARF1}\end{equation}
with initial conditions
\begin{equation}\begin{array}{c}
W(G,0)=1; \ \ \ \ W(s,0)=0, \ \ \ if \ s\ne G ,
\end{array}\label{PARF2}\end{equation}
where $\gamma_1$ and $\gamma_2$ are the constant rates of an elementary
coalescence act and of an elementary fragmentation act, respectively.
Equation (\ref{PARF1}) is constructed so as to absorb the conditions
$W(G+1,t\ge 0)=0, W(0,t\ge 0)=0$. We shall see below, the solutions of
Eq. (\ref{PARF1}) bear out this assertion.

It should be noted, that the mass spectra of clusters for process
Eqs. (\ref{PARF1}) and (\ref{PARF2}) differ from those in Sect.~\ref{MDPAP}, \ref{SCDPAP}.

Equation (\ref{PARF1}) can be solved for each $G$. If, for example, G=5,
\begin{equation}\begin{array}{c}
 W(s,t) = \sum_{i=1}^G C_i e^{\lambda_i t} A_{si} ,
\end{array}\label{PARF3}\end{equation}
{\footnotesize
\begin{equation}\begin{array}{c}
C =
[.44721, \,-.44721, \,-.44721, \,-.44721, \,
-.44721] ,
\end{array}\label{PARF4}\end{equation}
\begin{equation}\begin{array}{c}
\lambda  =
[-10.95389, \,-5.73117, \,-2.57163, \,
-.74329, \, 0] ,
\end{array}\label{PARF5}\end{equation}
} %end of \footn
{\footnotesize
\begin{equation}\begin{array}{c}
A=
 \left[
{\begin{array}{c}
.01096\,, \,.11376\,, \,-.43096\,, \,.77537\,, \,-.44721 \\
-.10909\,, \,-.53825\,, \,.67732\,, \,.19904\,, \,-.44721 \\
.42839\,, \,.67814\,, \,.36055\,, \,-.16309\,, \,-.44721 \\
-.77746\,, .19355\,, -.15969\,, -.36411\,,-.44721 \\
.44721\,, -.44721\,, -.44721\,, -.44721\,,-.44721
\end{array}}
 \right] .
\end{array}\label{PARF6}\end{equation}
}  % end of \footn

%\ins{fig4}{bbw.eps}{4in}{2in}%{h}
   \begin{figure}[h]
%    \centering
    \includegraphics[width=3in,totalheight=2in]{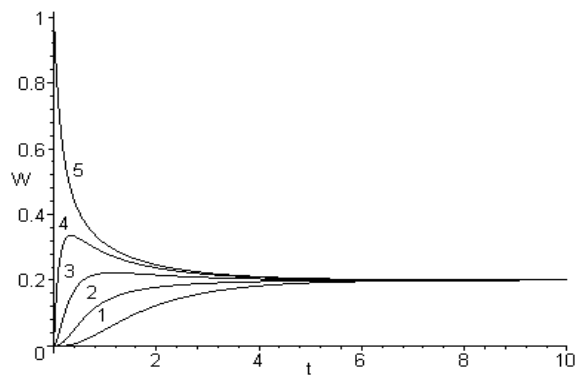}
    \caption{W-functions. Aggregation with an arbitrary fragmentation.  $G=5, \gamma=2, \gamma_1 =1, \gamma_2=1$.
    All curves $W(s,t)$ are marked by the number $s$ of the proper state.}
   \end{figure}

\section{\label{Res}Conclusion}

In this paper, we are interested in kinetics of cluster formation and
construct a number of analytically solved models of such a process. It does
not matter whether a motor lane, a computational network, space dust or glass
is a real environment to match such a script. The illustration of the problem
in terms of a traffic jam is not a specific one. The known technical and
natural phenomena of traffic jam or aggregation (coagulation) and their
evolution have been considered as processes of clustering (nucleation).

Brought forward in this work, the kinetics models take account of both the
aggregation (coagulation) and the fragmentation (decay) processes.
Previous studies of nonlinear Lotka--Volterra systems \cite{DGR} brought us
to a search for a possibility to give a linear description of those very
complicated and nonlinear situations or for something akin to such a picture.
A dynamical description of some system could be substituted by a stochastic
one. We consider one-dimensional cases (in the coordinate sense) but the
scenarios should be valid for the three dimensions if there is a spherical
symmetry.

When we deal with a system of a finite number of particles, a natural
way of the above substitution is to use the language of enumerated states of
that system. A state of that sort is characterized by the population number
and probability function to reveal the system itself in this state exactly.
The "three-level bunch" scheme of transitions between the three nearest states
of
the system is natural as well. Of course, that probability should depend upon
the probability of something else to happen. In the case considered, it could be
an act of coalescence or decay ( dissociation, fragmentation, etc., where the
term depends on the application ). Thus, we have just met a product of
probabilities. What can one do?

The probability (rate) of the above acts could be dependent or independent
on the system states or its particular attributes. In the case of such a
dependence, one can say nothing without a special investigation. On the
contrary, the rate independence of the above circumstances makes the situation
a linear one in the state probability function. Let us take into account that
the rates may be independent not only of macro parameters but also of the
cluster size (e.g., if
coalescence/fragmentation depends only on the valency of some chemical
clusters). This leads to our key mathematical assumption of the above rates
being constant. That is why we obtain linear analytically solved master
equations of that clustering kinetics, which are also evolutionary-type
equations.

The very method to solve most of those problems is in use of the generating
function.

But some peculiarities of applying of this method for solving the
linear differential-difference equations are revealed.
It becomes useless when the structure homogeneity of terms in the
r.h.s. of equations discussed is violated as a result of combining aggregation
and fragmentation terms, i.e., owing to violation of the Markovian semigroup
structure of those r.h.s. The latter, as well as symmetry, momenta and other
questions (e.g. universality of the three-level bunch scheme ) not included
in this sketch will be considered in more detail in papers to follow, as well
as some details of the utilization of a MAPLE-program, used to obtain some
analytical results, which is to be published in \cite{JCMSE}.

Other questions about the mass spectra and the single-cluster mass
distribution in a pure aggregation process help to display the combinatorial
nature of the problem under consideration and its closeness to some
population dynamics problem. They are solved by means of the Laplace
transformation, which leads to the calculation of certain convolutions. The
properties of those convolutions are true due to the isomorphism between a
set of generating functions with a product operation and a set of $w(g)$ with
a convolutions, which results in a useful formula.

It should be remembered, that we considered two types of fragmentation: the
similar fragmentation and the arbitrary one. They lead to different results.
Such a difference can be realized as a manifestation of distinction between
collective and more discrete additive properties of a cluster on a very
abstract level, far off a specific nature of that cluster. Moreover, we shall
see the possible existence of several aggregation mechanisms (for a variety
of reasons). Thus, one and the same stochastic description of the aggregation
can mask a number of processes diverse in their dynamical details.

However, the generality of the results is restricted. The conjecture that the
rates are constant ($\gamma_{1,2}=const_{1,2}$) is nowadays somewhat at odds with
what is known about molecular and atomic clusters, and processes for nanoscale
units of matter. Obviously, the mentioned conjecture reduces a mathematical
universality as well. We are grateful to the referee for drawing our attention
to the necessity of making this remark.

Nevertheless, the authors hope that appropriate processes of such a type may
be found by the following reason. The presented formalism is independent of
any physical scale. The only conjecture of the possibility to draw distinction
between certain particles and intervals between them has been done. Such an
admissibility could be combined with not only a classical conception but a
quantum one under some conditions (e.g., if the first Born approximation is
valid ) as well.

Present work belongs to the stream originated from the famous Smoluchovski`s
articles \cite{Smoluch}.

\begin{acknowledgments}
The authors are very grateful to V.B. Priezzhev for discussions
and numerous kind advice.
\end{acknowledgments}

\appendix*
\section{}

%\section{Appendixes}

Let
\begin{equation}\begin{array}{c}
 N_0={0,1,2,...},\\ \\
 a(n)=a_n,\ n\in N_0,\\ \\
 A=\{a:N_0\rightarrow R\},\\ \\
 F=\{f:C\rightarrow C,f(z)=\sum\limits_{n\in N_0}a_n z^n\}.
\end{array}\label{Append1}\end{equation}

The reversible mapping $Z:A\rightarrow F$
\begin{equation}\begin{array}{c}
 f=Z(a)\ \ so\ as\ \ f(z)=\sum\limits_{n\in N_0}a(n)z^n
\end{array}\label{Append2}\end{equation}
has the inverse mapping $Z^{-1}$
\begin{equation}\begin{array}{c}
 a=Z^{-1}(f)\ \ so\ as\ \ a(n)=\oint{f(z)\frac{dz}{z^{n+1}}},\\ \\
 n\in N_0.
\end{array}\label{Append3}\end{equation}

The convolution $\circ$ is an operation on a set A such that
\begin{equation}\begin{array}{c}
 c=a\circ b \ \ \ iff \ \ \  c(k)=\sum\limits_{j=0}^k a(k-j)b(j).
\end{array}\label{Append4}\end{equation}

The mapping Z is a morphism of a semigroup $(A,\circ)$ to a semigroup
$(F,\cdot)$.
Z maps the convolution $\circ$ to the product $\cdot$ :
\begin{equation}\begin{array}{c}
 c=a\circ b \rightarrow  Z(c)=Z(a)Z(b)
\end{array}\label{Append5}\end{equation}

The associative and commutative semigroup $(A,\circ)$ has the unit
$\Delta$:
\begin{equation}\begin{array}{c}
 \Delta(0)=1,\ \Delta(n)=0,\ n\ne 0,\\ \\
 Z(\Delta)=1,\ Z^{-1}(1)=\Delta.
\end{array}\label{Append6}\end{equation}

Here is a useful formula
\begin{equation}\begin{array}{c}
 (a_1 \circ a_2 \circ \ldots \circ a_n)(m) \\
 =\sum\limits_{\{j_k \in N_0\}} a_1(j_1)\ldots a_n(j_n)
 \Delta(m-j_1-\ldots -j_n)
\end{array}\label{Append7}\end{equation}

\end{document}